\documentclass{iaus}
\usepackage{graphicx}

\newcommand{\beq}{\begin{equation}}
\newcommand{\eeq}{\end{equation}}
\newcommand{\bea}{\begin{array}}
\newcommand{\eea}{\end{array}}
\newcommand{\alp}{\alpha}

\title[Migration and Location of Super Earths]
{Migration and Final Location of Hot Super Earths
in the Presence of Gas Giants}

\author[Zhou \& Lin ]
{Ji-Lin Zhou$^1$
 \and Douglas N.C. Lin$^{2,3}$}

\affiliation{$^1$ Department of Astronomy, Nanjing University,
Nanjing 210093, China \\ {email: zhoujl@nju.edu.cn } \\
$^2$UCO/Lick Observatory, University of California, Santa Cruz, CA
95064, USA \\ email: {\tt lin@ucolick.org } \\  $^3$Kavli Institute
of Astronomy and Astrophysics, Peking University, Beijing 100871,
China }

\pubyear{2008}
\volume{249}  
\pagerange{1-5}
 \jname{Exoplanets : Detection, Formation and
Dynamics} \editors{Y.-S. Sun, S. Ferraz-Mello  \& J.-L. Zhou, eds.}

\begin{document}

\maketitle

\begin{abstract}
Based on the conventional sequential-accretion paradigm, we have
proposed that, during the migration of first-born gas giants outside
the orbits of planetary embryos, super Earth planets will form
inside the 2:1 resonance location by  sweeping of mean motion
resonances (Zhou et al. 2005). In this paper, we study the
subsequent evolution of a super Earth $(m_1)$ under the effects of
tidal dissipation and perturbation from a first-born gas giant
($m_2$) in an outside orbit.  Secular perturbation and mean motion
resonances (especially $2:1$ and $5:2$ resonances) between $m_1$ and
$m_2$ excite the eccentricity of $m_1$, which causes the migration
of $m_1$ and results in a hot super Earth.  The calculated final
location of the hot super Earth is independent of the tidal energy
dissipation factor $Q'$. The study of migration history of a Hot
Super Earth is useful to reveal its $Q'$ value and to predict its
final location in the presence of one or more hot gas giants.  When
this investigation is applied to the GJ876 system, it correctly
reproduces the observed location of GJ876d around 0.02AU.

\end{abstract}

\firstsection 
\section{Introduction}

The search for habitable planets is an essential step in the quests to
unravel the origin of the Solar System and find life elsewhere.  To
date, more than 250 exoplanets are detected mainly by radial velocity
survey of nearby solar-type stars\footnote{http://vo.obspm.fr/exoplanetes/,
http://exoplanets.org/.}.  In the database, there are $17$ planets
with mass less than $25$ Earth mass ($M_{\oplus}$),
and among them 8 planets have orbits with period $<10$ days.
They are GJ876d, HD69830b, GJ674b, HD160691d, 55Cnc e, Gl581b,
HD219828b and GJ436b. We call them `hot super Earths'.

According to the conventional core-accretion scenario of planet
formation, planets form in a protoplanetary disk around the host
protostar. Through the sedimentation of dust, cohesive collisions of
planetesimals, many embryos will form by accreting and clearing the
planetesimals in their feed zone (a band centered on the embryo with a
width of $\sim 10$ Hill radius) and result in dynamically isolated
bodies.  In a disk with several ($f_d$) times of minimum mass solar
nebular, the isolation mass is (Zhou et al. 2007),
\beq
   M_{\rm iso}=0.51\times 10^{-2}M_\oplus \eta k_{\rm iso}^{3/2},
   \label{miso}
   \eeq
where
\beq
\begin{array}{l}
   \eta=(f_{\rm d}f_{\rm ice})^{3/2}(\frac{a}{\rm 1AU})^{3/4}
   (\frac{M_*}{M_\odot})^{-3/2}. \\
      \log(k_{\rm iso})=\sqrt{b^2+0.61c}-b, \\
    b=2.8+0.33\log \eta ,  \\
    c=3.6+0.67\log \eta+\log T_{\rm dep},\\
   \end{array}
\eeq
$f_{\rm ice}=1$ for embryos inside  the ice line and  $f_{\rm ice}=4.2$ outside that,
$T_{\rm dep}$ is the timescale of depletion of gas disk.

According to equation (\ref{miso}), the isolation mass inside the ice line
   $(a\approx  2.7 $AU in solar system) is too small to become
a super Earth.  Unless their nascent protostellar disks
are highly compact, the observed super Earths are unlikely formed in
situ.  Some extra mechanisms are required to account for the excitation of
eccentricity and the merge of isolated embryos into super Earths.  In
Zhou et al. (2005), we have proposed two mechanisms that may lead to the
excitation of eccentricities of embryos: (1) During the type-II
migration of a first-born gas giant planet outside the orbits of
embryos, the locations of its mean motion resonances (mainly 2:1
resonance) sweep through the embryos region; (2) During the dispersal
of the gas disk, the location of secular resonance between the gas
giant and embryos sweeps through the inner orbits.  Additional
mechanisms have also been discussed by Raymond et al. (2007).

In this paper, we suppose a super Earth has formed through one of
the above mechanisms, and study the subsequent evolution after the
gas disk was depleted and the gas giant has stopped its migration.
First we briefly review the secular evolution of two planets under
tidal dissipation. Then we show some numerical results in section 3.
Conclusions are presented in section 4.

\section{Secular dynamics under tidal dissipation}

\subsection{Tidal perturbation timescale}
We adopt a two-planet system as a model.  Suppose two planets with
mass $m_i(i=1,2)$ (in the order form inner to outer) moving around a
star with mass $m_*$ in the same orbital plane. Let $m_1$ be an
Earth-like planet, and $m_2$ a gas giant, $S_i,\Omega_i, a_i,r_i$
are the radius, spin rate (with spin axis perpendicular to the
orbital plane), semi major axis, distance from the star of planet $i
~(i=1,2 {\rm respectively})$, . The acceleration to the relative
motion of $m_i$ caused by the tidal interaction between the star and
planet $m_i$ has the form of (Mignard 1979, Mardling \& Lin 2002)
 \beq {\bf F}_{\rm
i,tid}=-(1+\lambda^{-1})\frac{9n_i}{2Q'_i}(\frac{m_*}{m_i})(\frac{S_i}{a_i})^5(\frac{a_i}{r_i})^8 [3
v_{ir} \hat{r}+ (v_{i\phi} - r_i \Omega_i) \hat \phi],
  \label{ftid}
\eeq where $\hat{r}, \hat \phi$ are the unit vector of radial and
transversal direction of the orbital plane, ${\bf V_i}= v_{ir}
\hat{r}+v_{i\phi} \hat \phi$ and $n_i$ are the Kepler velocity and
mean motion of planet $i$(i=1,2), respectively,  $Q'_*$ and $Q'_i$
are the effective tidal dissipation factor of the star and planet
$i$ defined as $Q'=3Q/(2k_L)$, where $Q^{-1}=\tan (2 \epsilon)$ is
the effective dissipation function, $\epsilon$ is the tidal lag
angle (Goldreich \& Soter 1966), $k_L$ is the Love number or twice
the apsidal constance for gaseous planets(e.g., Mardling \& Lin 2002), and
 \beq
 \lambda=(\frac{Q'_*}{Q'_i})(\frac{m_*}{m_i})^2(\frac{S_i}{S_*})^5
 \label{lambd}
\eeq
is the ratio of  tidal dissipation in the planet  to that in the star.
If $\lambda \gg 1$, tidal dissipation in the planet dominates the evolution.

The values $Q'_*$ inferred form the observation of circularization period in various stellar clusters
are $\sim 1.5 \times 10^5 $ for young stars with age less than $0.1$Gyr, and $\sim 10^6 $ for mature stars (Terquem et al. 1998,Dobbs-Dixon et al. 2004).
The $Q'$ value for Jupiter inferred  form Io's orbit evolution ranges from $5\times 10^4$ to $2\times 10^6$ (Yoder \& Peale 1981).
And for Earth, $Q'_E\approx 60$ (Yoder 1995).
Thus for a gas giant planet with Jupiter mass, suppose $Q'_* \approx Q'_J=10^5$, from Eq.(\ref{lambd}), $\lambda \sim 10$, while
for a terrestrial planet with Earth mass, $\lambda\sim 10^4$. So in the case of tidal interaction between
an Earth-like planet and a star,  tidal dissipation in the planet dominates the evolution,
thus we neglect the contribution of tide in star in the following study.

Take $m_1$ as an example. Under the perturbation of tidal effect, the averaged equations (over a period of
orbital motion) governing the  evolution of  planet $m_1$ are,
\beq
\bea{l}
<\dot{a}_1>_{\rm tide}=-2 a_1\tau_{\rm tide}^{-1}  \left[f_1(e_1)
- (\frac{\Omega_1}{n_1})f_2(e_1)  \right], \\
<\dot{e}_1>_{\rm tide}=-9  e_1 \tau_{\rm tide}^{-1} \left[f_3(e_1) -
\frac{11}{18}(\frac{\Omega_1}{n_1})f_4(e_1)\right], \\
<\dot{\varpi}_1>_{\rm tide}=<\dot{\lambda}_1>_{\rm tide}=0.
\eea
\label{dotae}
\eeq
where $\varpi_1,\lambda_1$ are the longitude of perihelion and
 mean longitude of the orbit of $m_1$ (with volume density $\rho_1$), respectively, and
\beq \tau_{\rm tide}=\frac{4Q'_1}{63n_1}(\frac{m_1}{m_*})(\frac{a}{S_1})^5=
2.4\times 10^7  Q'_1
  (\frac{a_1}{\rm 0.1AU})^{\frac{13}{2}}
  (\frac{m_*}{m_\odot})^{-\frac{3}{2}}
  (\frac{m_1}{m_\oplus})^{-\frac{2}{3}}
   (\frac{\rho_1}{\rm 3 g~cm^{-3}})^{\frac{5}{3}}~ {\rm yr}.
\eeq
Functions used are:
\beq
\begin{array}{l}
 f_1(e)=(1+\frac{31}{2}e^2+\frac{255}{8}e^4+\frac{185}{16}e^6
 +\frac{25}{64}e^8)/(1-e^2)^{15/2}, \\
 f_2(e)=(1+\frac{15}{2}e^2+\frac{45}{8}e^4+\frac{5}{16}e^6)/(1-e^2)^{6}, \\
 f_3(e)=(1+\frac{15}{4}e^2+\frac{15}{8}e^4+\frac{5}{64}e^6)/(1-e^2)^{13/2},\\
f_4(e)=(1+\frac{3}{2}e^2+\frac{1}{8}e^4)/(1-e^2)^{5}, \\
f_5(e)=(1+3e^2+\frac{3}{8}e^4)/(1-e^2)^{9/2}, \\
f_6(e)= (1+\frac{15}{7}e^2+\frac{67}{14}e^4+\frac{85}{32}e^6+
\frac{255}{448}e^8+\frac{25}{1792}e^{10})/(1+3e^2+\frac38 e^4),\\
f_7(e)=(1+\frac{45}{14}e^2+8e^4+\frac{685}{224}e^6+
\frac{255}{448}e^8+\frac{25}{1792}e^{10})/(1+3e^2+\frac38 e^4).
\end{array}
\eeq
The evolution of spin rate  $\Omega_1$ is subjected to,
\beq
   I_1 \dot{\bf \Omega}_1= -\frac{m_*m_1}{m_*+m_1} {\bf r}_1 \times {\bf F}_{\rm 1, tide}
\eeq
where $I_1 \approx  \frac25 m_1 S_1^2$ is the inertial momentum of
$m_1$.
The averaged change rate is
\beq
<\dot{\Omega}_1>_{\rm tide}= \frac52 \tau_{\rm tide}^{-1} (\frac{a_1}{S_1})^2   \left[f_2(e_1) -
(\frac{\Omega_1}{n_1})f_5(e_1) \right].
\eeq
A stable equilibrium configuration occurs at
\beq
 \Omega_{1,eq}=\frac{f_2(e_1)}{f_5(e_1)}n_1.
\label{eql}
\eeq
 Since the timescale to reach the
equilibrium state ($\sim  \tau_{\rm tide} (S_1/a_1)^2$)  is several orders less than the tidal circularization
 timescale, we suppose such a state is reached. Substitute
Eq.(\ref{eql}) into (\ref{dotae}),  we derive the timescales of tidal evolution of $m_1$,
\beq \tau_{\rm a-tide}\equiv \frac{a_1}{\dot{a}_{\rm 1}}=-
\frac{(1-e_1^2)^{15/2}}{ 2 e_1^2f_6(e_1)}\tau_{\rm tide} , ~~~
\tau_{\rm e-tide} \equiv \frac{e_1}{\dot{e}_{\rm 1}}=
-\frac{(1-e_1^2)^{13/2}}{  f_7(e_1) }\tau_{\rm tide}.
\label{aetidt} \eeq
 Note that,   $\tau_{a-tide} \gg
\tau_{e-tide}$ when $e_1 \approx 0$. However, when
 $e_1 \approx 1$, $\tau_{a-tide}$ and $ \tau_{e-tide}$ could be very small,
and $\tau_{a-tide} < \tau_{e-tide}$ as long as $e> 0.63425...$.

Due to the huge difference of
$Q'$ between the Earth-like planet $m_1$ and the gas giant $m_2$, for
our later investigation of tidal evolution with $a_1<0.63a_2$,  we
 neglect the tidal effect in planet  $m_2$.

\subsection{Secular evolution in the case of $e_1\ll e_2$}

When  $e_1\ll e_2$, the secular evolution of $m_1,m_2$ under
tidal dissipation and general relativity effect can be approximated by
the following equations (Mardling 2006):

\beq
\begin{array}{l}
\dot{e}_1=-W_o e_2 \sin \eta -W_T e_1, \\
\dot{e}_2=W_ce_1\sin \eta, \\
\dot{\eta}=W_q-W_o (\frac{e_2}{e_1})\cos \eta,
\end{array}
\eeq
where $\eta=\varpi_1-\varpi_2, \alp=a_1/a_2,\beta=\sqrt{1-e_2^2}$, and
\beq
\begin{array}{l}
W_o=\frac{15}{16} n_1 (\frac{m_2}{m_*}) \alpha^{4} \beta^{-5}, \\
W_T=\tau_{e-tide}^{-1},~~W_c=\frac{15}{16} n_2 (\frac{m_1}{m_*}) \alpha^{3} \beta^{-4} ,\\
W_q=\frac{3}{4} n_1 (\frac{m_2}{m_*}) \alpha^{3} \beta^{-3} [1-\sqrt{\alpha} (\frac{m_1}{m_2})\beta^{-1} +\gamma \beta^3],
\end{array}
\eeq
with $\gamma=4(n_1a_1/c)^2(m_*/m_2)\alp^3$, the ratio of general relativity to quadruple contribution of $\dot{\eta}$.
According to these equations, the secular evolution of $e_1$ and $e_2$
 mainly passes three stages:
\begin{description}
   \item[(1)]~After a short time oscillation,
    the evolution of $e_1$ and $\eta$ reaches  a state of
    librating around a quasi-equilibrium configuration
   with $e_1=e_1^{eq}$ and $\eta=2n\pi$ or $(2n+1) \pi$ , where (Mardling 2006)
\beq
   e_1^{eq}=e_2\frac{ W_0}{|W_q|}=
   \frac{5/4\alp e_2 }{\beta^2|1-\sqrt{\alp}(m_1/m_2)\beta^{-1}+\gamma \beta^3|}.
   \label{e1eq}
\eeq
   \item[(2)]~As $\eta$ librates and $e_1$ evolves to
   $e_1=0$ gradually,
   $a_1$ is damped according to Eq.(\ref{aetidt}), thus  $m_1$ migrates inward
   efficiently.
     \item[(3)]~Finally $e_2$ is damped on a timescale $\tau_c \gg \tau_{e-tide}$. 
    During this timescale,   the
    orbit of $m_1$ is almost  circularized, and
    migration of $m_1$ is effectively stopped at a location $a_{1f}$.
 \end{description}

The location of $a_{1f}$ is what we want to find. However, due to the
presence of resonant motion, the evolution of the two-planet system
in real situation is more complicated, as we will show below.

\begin{figure}[b]
\vspace*{1.5 cm}
\begin{center}
\includegraphics[width=3.8in]{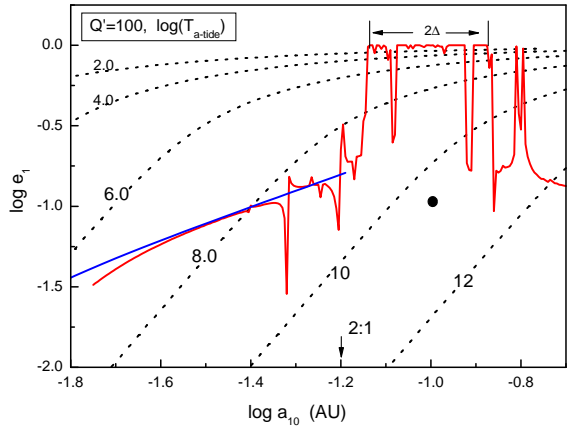}
\vspace*{-3.2 cm}
 \caption{Maximum eccentricity (red solid line ) of $m_1$  in an initial circular orbit
 of semi-major axis $a_{10}$  excited by $m_2$ (black circle, with $a_{20}=0.1$AU,
$e_{20}=0.1$). The black  dotted lines with labels $2.0,4.0,...$
denote  the timescale ($\log(T_{a-tide}/{\rm years})$)  of $m_1$
from Eq. (\ref{aetidt}) at the specific location of ($a_{10},e_{1}$) with $Q'_1=100$.
The blue dashed line
is obtained by two times of
the equilibrium values defined by Eq.(\ref{e1eq}).  }
   \label{fig1}
\end{center}
\end{figure}

\section{Numerical simulations}
\label{model}

\begin{figure}[b]
\vspace*{3.0 cm}
\begin{center}
\includegraphics[width=4.8in]{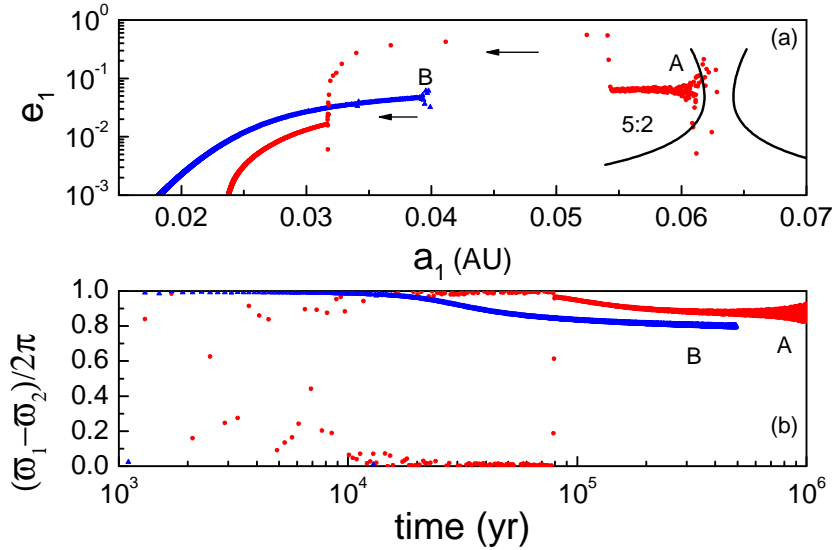}
\vspace*{-4.0 cm}
 \caption{Evolution of orbits with $a_{10}=0.063$ (orbit A) and $a_{10}=0.040$ (orbit B).  (a) Evolution track in $a_1-e_1$ plane.
 The dashed  line shows the width of $2:1$ resonance obtained from the circular restricted-three-body problem. The $5:2$ indicates the
 $5:2$ resonance location at the place that the eccentricity of orbits A jumps up. The arrows indicate the  evolution directions.
 (b) Evolution of $(\varpi_1-\varpi_2)$ (in radian).  }
   \label{fig2}
\end{center}
\end{figure}

\begin{figure}[b]
\vspace*{3.0 cm}
\begin{center}
\includegraphics[width=4.8in]{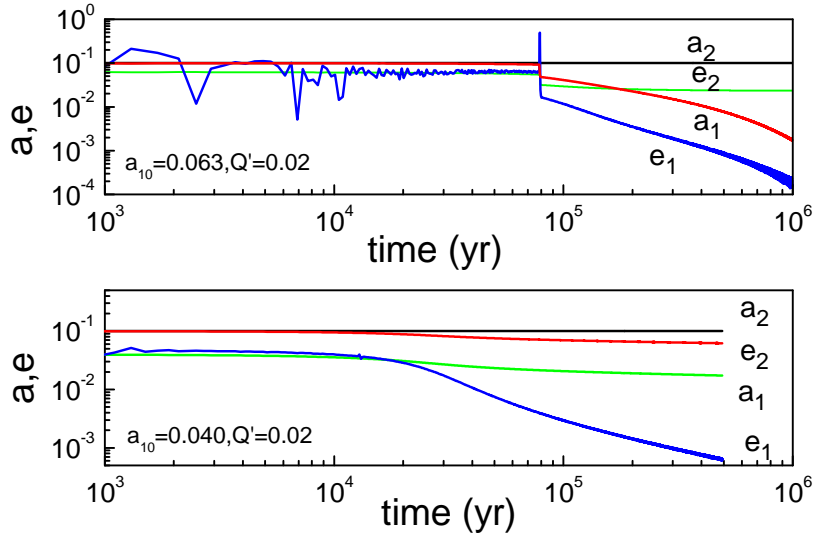}
\vspace*{-4.0 cm}
 \caption{Evolution of orbital elements for the two orbits shown in Fig.2 with
 initial elements $a_{10}=0.040,0.063$, respectively. The evolution of orbit with $a_{10}=0.040$
 fits with  the secular evolution described in section 2, but for the orbit
 with $a_{10}=0.063$, the presence of the resonance leads to a dramatic
 increase (decrease) of $e_1$($e_2$) at time $t \approx 8\times 10^4 $ year,
 when the orbit crosses the 5:2 resonance in Fig.2.}
   \label{fig3}
\end{center}
\end{figure}

\begin{figure}[t]
\vspace*{3.0 cm}
\begin{center}
\includegraphics[width=4.0in]{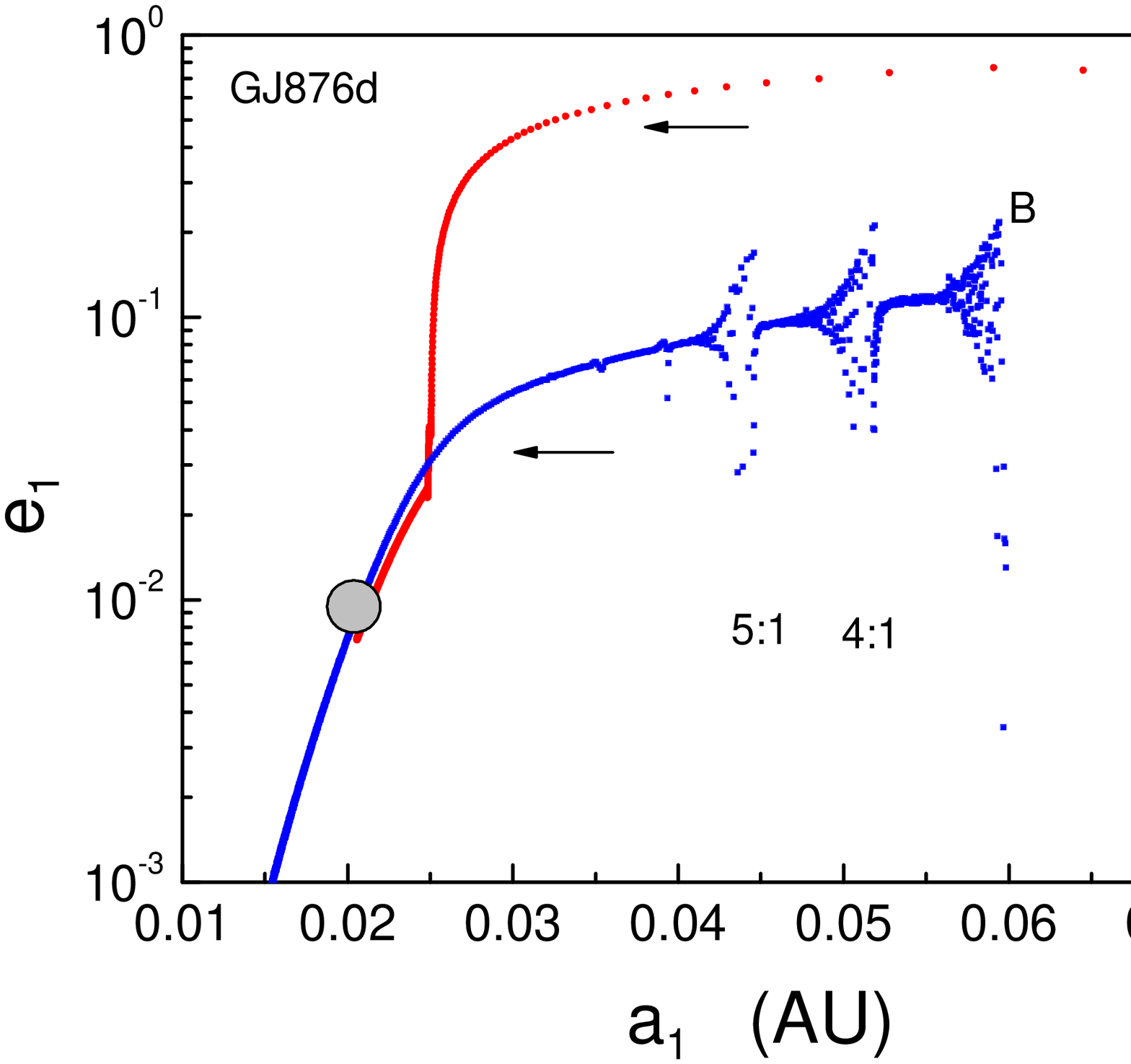}
\vspace*{-3.5 cm}
 \caption{ Evolution track of orbits with $a_{10}=0.07074$
  (the location of $5:2$ resonance, orbit A) and $a_{10}=0.060$ (orbit B).
  The arrows indicate the  evolution directions.}
   \label{fig4}
\end{center}
\end{figure}

We study the migration of an Earth-like planet under the tidal and
mutual planetary perturbations with general three-body
model. The system consists of a solar mass host
star $(m_*=1M_\odot)$, an Earth-like planet with mass
($m_1=5M_\oplus$), and a Jupiter mass gas giant ($m_2=M_J$, where
$M_J$ is Jupiter mass).
Let $m_1$ be initially in a nearly circular orbit ($e_{10}=10^{-3})$, and
$m_2$ with initial elements $a_{20}=0.1$AU,$e_{20}=0.1$.  To shorten
the integration time, we let $Q'=0.02$, as the migration timescale is
proportional to $Q'$.

By integration of the full equations of the general three-body system
without tidal dissipation, we plot the maximum eccentricity ($e_{\rm
1max}$) of $m_1$ excited by $m_2$ in Figure 1.  The corresponding
tidal-damping timescale obtained from equation (\ref{aetidt}) with
$(a_{10}, e_1)$ is also shown in the background of Fig.1.

As we can see from Figure 1, the orbits of $m_1$ at most locations with
$a_{10}<0.16 $AU have $T_{\rm a-tide}< 10$Gyr.  However,
most of those orbits in the Hill unstable region around $m_2$ with half width
 $\Delta=(e_{20}+2\sqrt{3}h) a_2 \approx 0.034$AU (where
$h=[m_2/(3m_*)]^{1/3}$) will be scattered to far away in our coplanar model.
  According to Zhou et
al. 2005, embryos formed inside the location of the $2:1$
resonance($a_{2:1}\approx 0.063$AU)
with $m_2$  are dynamically stable, so we
focus on the evolution of orbits with initial semi major axis $a_{10}
\le 0.063$AU.

If planet $m_1$ is initially located in lower order mean motion resonances with
$m_2$, its  eccentricity will be excited, thus a fast inward migration
of $m_1$ is induced,  according to Eq. (\ref{aetidt}).
 Fig. 2 shows the evolution of two orbits
either from $2:1$ resonance location ($a_{10}=0.063$AU) or from
non-resonance location ($a_{10}=0.040$AU).
 During the subsequent passage through $5:2$
resonance, the amplitude of eccentricity excitation is relatively large.
Recall that, according to Eq. (\ref{aetidt}), the $a-$damping
timescale is much smaller than that of $e$-damping at high
eccentricity.  Thus a fast migration occurs until $e_1$ decrease
to a small value $\sim 0.01$ (Novak {\it et al.} 2003).  Then a slow
migration linked with the secular dynamics occurs, with
$\eta=\varpi_1-\varpi_2$ librating along an equilibrium value(see
Fig.2b).

The migration induced by resonant eccentricity-excitation is different
from  that excited by mutual secular perturbation. When we check the
evolution of $a_i,e_i, (i=1,2)$ during the passage of $5:2$ resonance,
we find that $a_1,e_1,e_2$ have dramatic decrease after the crossing
the  $5:2$ resonance (Fig.3). The decrease of $e_2$ causes
the different final states (i.e., the final state of orbit A and B) in
$(a_1,e_1)$ plane of Fig.2.  The final location is around $0.018\sim
0.025$AU, depending on the different evolutionary routines.

In order to show that the above track correctly reproduces the observed
location of extrasolar planets, we applied this study to the GJ876
system.  GJ876 is a M dwarf star located 4.72 pc away from us in the
solar neighborhood.  To date, two gas giant planets, GJ876b and GJ876c,
were observed to be located on orbits with period around $30$ days and $60$
days, an example of 2:1 mean motion resonance, and a hot planet GJ876d
with mass around $5.7 M_{\oplus}$ in an orbit with period 1.94
days($a=0.0208$AU, e=0).  We numerically simulate the evolution of an
Earth-like planet inside a gas giant located in the present orbit
($a_{20}=0.13$AU, $e_{20}=0.2243$). Figure 4 shows the evolution track
of $m_1$. The final location is around $0.02$ AU according to the
simulation, which is almost independent of the initial location of
$m_1$.

\section{Summary and discussions}

Many super Earths are observed to be located inside the orbits of gas
giants. These super Earth and gas giant pairs may be a natural
consequence of planet formation and migration. Embryos formed prior and
interior to the gas giants are induced to migrate, collide, and evolve
into close-in Super Earths(Zhou et al. 2005). In this report, we have shown that, the
migration of super Earths under tidal dissipation and the perturbation
from gas giants is mainly along the secular evolution paths.  Although
resonances between super Earths and gas giants may excite the
eccentricity and speed the migration timescale, the final evolution
path can be well determined.

According to the investigation of this paper, we find that the study of
the evolution path provides useful information in the following
ways: (i) the migration path shows the evolution history, especially
the evolution of $e_1,e_2$, $a_1$ (Figs.2,4). (ii) Comparing  the observed
location of the planet in the path, we can deduce the range of $Q'$ values for
the hot Super Earths. We will investigate in more details on these topic
in the future.

\acknowledgments This work is supported by NSFC(10778603,10233020),
National Basic Research Program of China(2007CB814800), NASA
(NAGS5-11779, NNG04G-191G, NNG06-GH45G), JPL (1270927),
NSF(AST-0507424, PHY99-0794).

\end{document}